\documentclass{elsart}
\usepackage{graphicx}

\begin{document}

\def\simlt{\mathrel{\rlap{\lower 3pt\hbox{$\sim$}}\raise 2.0pt\hbox{$<$}}}
\def\simgt{\mathrel{\rlap{\lower 3pt\hbox{$\sim$}} \raise 2.0pt\hbox{$>$}}}
\def\di{\mbox{d}}
\def\Msun{M_{\odot}}
\def\HI{\hbox{H$\scriptstyle\rm I\ $}}
\def\cm3{\;\mbox{cm}^{-3}}
\def\SNyy{\mbox{SN}_{\gamma\gamma}}
\def\etal{{\it et al.~}}

\newcommand{\q}{\begin{equation}}
\newcommand{\qa}{\begin{eqnarray}}
\newcommand{\qs}{\begin{eqnarray*}}
\newcommand{\nq}{\end{equation}}
\newcommand{\nqa}{\end{eqnarray}}
\newcommand{\nqs}{\end{eqnarray*}}

\begin{frontmatter} 
\title{Induced Formation of Primordial Low-Mass Stars}
\author[SISSA]{R. Salvaterra},
\author[SISSA]{A. Ferrara}, 
\author[FI1,FI2]{R. Schneider}

\address[SISSA]{SISSA/International School for Advanced Studies, Via Beirut 4, 34100 Trieste, Italy}
\address[FI1]{INAF/Osservatorio Astrofisico di Arcetri, L.go E. Fermi 5, 50125 Firenze, Italy}
\address[FI2]{''Enrico Fermi'' Center, Via Panisperna 89/A, 00184 Roma, Italy}
 
\maketitle \vspace {7cm }
 
\begin{abstract}
We show that the explosion of the first supernovae can trigger 
low-mass star formation via gravitational fragmentation of the supernova-driven 
gas shell.  If the shell mass does not exceed the host galaxy gas mass,
all explosions with energies $E_{SN} \geq 10^{51}$~erg
can lead to shell fragmentation. However, the minimum ambient density required
to induce such fragmentation is much larger, $n_0 > 300\;\cm3$, for Type II supernovae
than for pair-instability ones, which can induce star formation 
even at lower ambient densities. 
The typical mass of the unstable fragments is
$\sim 10^{4-7}\,\Msun$; their density is in the range $110-6\times10^7\;\cm3$.
Fragments have a metallicity strictly lower than $10^{-2.6}\,Z_\odot$ and
large values of the gravitational-to-pressure force ratio 
$f\simeq 8$. Based on these findings, we conclude that the second generation of 
stars produced by such self-propagating star formation is predominantly 
constituted by  low-mass, long-living, extremely metal-poor (or even metal-free, 
if mixing is suppressed) stars. We discuss the implications of such results for 
Pop III star formation scenarios and for the most iron-poor halo star HE0107-5240.
\end{abstract}

\begin{keyword}
galaxies: formation - first stars - cosmology: theory;\\

98.80\sep 26.30 \sep 98.65.D \sep 97.10.B
\end{keyword}
 
\end{frontmatter}

\section{Introduction}

As many recent numerical (Bromm, Coppi \& Larson 1999, 2002; Abel, Bryan 
\& Norman 2000) and semi-analytical (Schneider \etal 
2002, Omukai \& Inutsuka 2002, Omukai \& Palla 2003) studies have
shown, the first luminous objects, the so called Population III (Pop III)
stars, are likely to be very massive. According to these studies, stars
with a characteristic mass of 100-600 $\Msun$ originate in the primordial gas, 
whereas low mass objects should not have formed due to the lack of heavy
elements. High mass stars are indeed required to account for the unexplained
Near Infrared Background (Salvaterra \& Ferrara 2003; Magliocchetti, 
Salvaterra \& Ferrara 2003).
Bromm et al. (2001) and Schneider et al. (2002) have shown that there exists a critical 
metallicity ($Z_{cr}=10^{-5\pm 1}\,Z_\odot$) setting the transition from a 
high-mass to a low-mass fragmentation mode of star formation. The exact value 
of $Z_{cr}$ depends on the fraction of heavy elements that are depleted onto 
dust grains (Schneider et al. 2003). In this scenario, a gas cloud with a 
metallicity $Z\sim 10^{-5.1}\,Z_\odot$ can lead to the formation of low mass 
stars if $\sim 20$\%
of the heavy elements is in dust grains (Schneider et al. 2003). 
The recent discovery of the most iron-poor star ([Fe/H]$=-5.3\pm0.2$) ever seen in 
our Galaxy (Christlieb et al. 2002) could be an example of such low-mass
metal poor stars originated from a gas with mean metallicity of 
$10^{-5.1}\,Z_\odot$ (Schneider et al. 2003; but see Umeda \& Nomoto (2003), 
who have pointed out that this star could have instead formed from a 
carbon-rich, iron-poor gas, with a corresponding  mean metallicity of 
$10^{-2}\,Z_\odot$). 

An alternative potentially viable mechanism
to form low-mass, metal-poor stars does exist. Indeed, 
Nakamura \& Umemura (2001, 2002, hereafter NU01 and NU02) 
have studied the collapse and fragmentation
of filamentary primordial gas clouds using one- and two-dimensional 
hydrodynamical simulations coupled with the nonequilibrium processes of
molecular hydrogen formation. They have shown that filaments with relatively
low initial density ($n\simlt \mbox{few}\times 10^{4}\;\cm3$) tend to fragment 
into dense clumps before the central density reaches $10^8-10^9\;\cm3$;
the fragment mass is around 100 $\Msun$. In contrast, if a filament has
initially a larger density, fragmentation continues until the clumps 
become optically thick to H$_2$ lines, leading to a typical fragment mass of 
$\sim 1\;\Msun$. So the resulting IMF of first stars would be bimodal, 
with a  low-mass peak around $\sim 1-2\;\Msun$ and a high-mass peak at a few 
hundred $\Msun$. Mackey et al. (2003) have suggested
that these low-mass peak stars could form in the gas shocked by the explosion
of the first generation of very massive supernovae (SNe).

In this paper, we put this hypothesis on a more quantitative basis. 
We will show that the shell is gravitationally unstable only for large 
explosion energies and that gravitationally unstable 
clumps are likely to form long-living, low-mass, 
metal-poor stars, that could be detected through  dedicated surveys as the
Hamburg/ESO objective prism survey (Christlieb et al. 2001).

The paper is organized as follows. In Section 2 we calculate the SN-driven
shell evolution and derive the conditions for gravitational instability.
In Section 3 we examine the instability
in the expanding shell for different explosion energies and interstellar
medium (ISM) densities,
whereas in Section 4 we calculated the properties of unstable fragments. 
In Section 5 we constrain the fragment metallicity; 
Section 6 discusses the
consequences for Pop III star formation scenarios.

\section{Evolution of the shell}

One can show that as far as the ambient gas pressure and cooling can be 
neglected, the shell evolution is described 
by the analytic expression derived by Sedov (1959):

\q
R_{sh}\propto \left(\frac{E_{SN}}{\rho_0}\right)^{1/5} t^{2/5},
\nq

\noindent
where $t$ is the time elapsed from the explosion, $E_{SN}$ is the total 
explosion energy, and $\rho_0=\mu m_H n_0$ is the density of the ISM
($n_0$ is the gas number density and $\mu=0.59$ is the mean molecular weight
of a primordial ionized H/He mixture).


The further evolution of a SN-driven shell in the interstellar medium can be
studied using the thin shell approximation 
(Ostriker \& McKee 1988; Madau, Ferrara \& Rees 2001). The momentum and 
energy conservation yield the following relevant equations:

\q\label{eq:mom}
\frac{d}{dt}(V_{sh}\rho_0 \dot{R}_{sh})=4\pi R_{sh}^2(P_b-P),
\nq
\q\label{eq:energy}
\frac{dE_b}{dt}=-4\pi R_{sh}^2P_b\dot{R}_{sh} - V_{sh}\bar{n}^2_{H,b}\Lambda(\bar{T}_b),
\nq

\noindent
where the subscripts `{\it sh}' and `{\it b}' indicate shell and bubble 
quantities, respectively. Here, $R_{sh}$ is the shell radius, $V_{sh}=(4\pi/3)R_{sh}^3$ 
is the volume enclosed by the shell; the overdots represent time derivatives. 
$P$ is the ISM pressure. The shell expansion is driven by the 
internal energy $E_b$ of the hot bubble gas, whose pressure is 
$P_b=E_b/2\pi R_{sh}^3$ (for a gas with adiabatic index $\gamma=5/3$). Finally,
$\bar{n}^2_{H,b}\Lambda(\bar{T}_b)$ is the cooling rate per unit volume
of the hot bubble gas, whose average hydrogen density and temperature are
$\bar{n}_{H,b}$ and $\bar{T}_b$, respectively. 
 The right-hand side of eq.
(\ref{eq:mom}) represents the momentum gained by the shell from the SN-shocked
wind, while the right-hand side of eq. (\ref{eq:energy}) describes the 
mechanical energy input, the work done against the shell, and the energy 
losses due to radiation.

As cooling becomes important for the swept up gas, the shock is no
longer driven by the heated gas and the evolution enters the momentum 
conserving phase, satisfying the familiar `Oort snowplow' solution, 
$R_{sh}\propto t^{1/4}$.

\subsection{Cooling time}

The cooling time is given by 

\q
t_{cool}=\frac{3}{2}\frac{kT_{ps}}{n_{ps}\Lambda(T_{ps})},
\nq

\noindent
where $n_{ps}= n_0 (\gamma+1)/(\gamma-1)$ is the post-shock density and
$T_{ps}$ the post-shock temperature for an adiabatic shock,  and  
$\Lambda(T_{ps})$ is the cooling function for a primordial gas. 
The evolution of
the bubble gas temperature $T_b$ is given by (Madau, Ferrara, Rees 2001) 

\q\label{eq:dTb}
\frac{dT_b}{dt}=3\frac{T_b}{R_{sh}}\dot{R}_{sh}+\frac{T_b}{P_{sh}}\dot{P}_{sh} -
\frac{23}{10}\frac{C_1}{C_2}\frac{kT_b^{9/2}}{R_{sh}^2P_{sh}}.
\nq

\noindent
where $C_1=16\pi\mu m_p\eta/25k$ and $C_2=(125/39)\pi\mu m_p$ 
and $\eta=6\times10^{-7}$ (c.g.s. units) 
is the classical Spitzer thermal conduction coefficient 
(we have assumed a Coulomb logarithm equal to 30). 
This relation closes the system of equations (\ref{eq:mom})-(\ref{eq:energy}).

If the shell lives for sufficiently long time (i.e. $t_{cool}\ll t$), 
the gas cools down to a final temperature $T_{f}\sim300$ K. This temperature
is the minimum temperature provided by cooling from H$_2$ molecules forming 
under nonequilibrium conditions in the post-shock gas (Ferrara 1998).

\subsection{Perturbations in the shell}

The problem of the fragmentation of expanding shell has been adressed by
several authors (Ostriker \& Cowie 1981; Elmegreen 1994; W\"{u}nsch \& Palous 2001;
Ehlerova \& Palous 2002). Here we apply these studies to the case of 
first SNe exploding in a medium of primordial composition.

The linear growth of perturbations in an expanding shell was derived
by Elmegreen (1994). Expansion tends to suppress (stabilize) the growth 
of  density perturbations owing to stretching, hence counteracting the
self-gravity pull. The instantaneous maximum growth rate is

\q\label{eq:w}
\omega=-\frac{3\dot{R}_{sh}}{R_{sh}}+\left[\left(\frac{\dot{R}_{sh}}{R_{sh}}\right)^2+\left(\frac{\pi G\rho_0 R_{sh}}{3c_{s,sh}}\right)^2\right]^{1/2},
\nq

\noindent
where $c_{s,sh}=(kT_f/\mu m_H)^{1/2}$ is the sound speed in the shell. 
Instability occurs only if $\omega>0$, or 

\q\label{eq:w2}
\frac{\dot{R}_{sh}}{R_{sh}}<\frac{1}{8^{1/2}}\frac{\pi G\rho_0 R_{sh}}{3c_{s,sh}}\propto \frac{t_{cross}}{t_{ff}^2},
\nq

\noindent
where $t_{cross}\sim R_{sh}/c_{s,sh}$ is the crossing time in the shell and $t_{ff}$ 
is the free-fall time. Furthermore, if the fragment mass is close to (but
slightly above) the Jeans 
mass, the $\omega >0$ condition translates into  $\dot{R}_{sh}/R_{sh}<1/t_{ff}$.
So, large shell velocity-to-radius ratios inhibit the formation of 
gravitationally unstable fragments.
These relations are derived under the assumption of supersonic motion, i.e.
$\dot{R}_{sh}(t)>c_{s,0}$, where $c_{s,0}$ is the sound speed in the ISM. If the 
shock decays into a pressure wave before the onset of the instability 
(i.e. before $\omega>0$), no fragmentation will take place and the shell 
will be dispersed by random gas motions. 

\bigskip

The most rapidly growing mode has wavelength

\q\label{eq:lfrag}
\lambda_f=\frac{6c_{s,sh}^2}{G\rho_0 R_{sh}}[1+(1-\xi^2)^{1/2}]^{-1},
\nq


\noindent
where the dimensionless parameter $\xi$ is given by

\qs
\xi=\left(\frac{8^{1/2}\dot{R}_{sh}}{R_{sh}}\right)\left(\frac{\pi G\rho_0 R_{sh}}{3c_{s,sh}}\right). \nonumber
\nqs

\medskip
\noindent
We will show later that the typical fragment size ($R_f=\lambda_f/2$) is 
greater than the thickness of the shell. In the radiative phase the thickness
is determined by the so-called cooling length, i.e. the distance travelled by 
the gas accelerated at the postshock velocity in a cooling time: 
$\Delta R_{sh} \approx \dot{R}_{sh} t_{cool}$. In this case
it is likely that disk-like fragments form with mass

\q\label{eq:mfrag}
M_f=\pi\rho_{sh}\Delta R_{sh} R_f^2,
\nq

\noindent
where $\rho_{sh}$ is the mean density in the thin shell, given by the ratio 
between its mass $M_{sh}$ (equal to the shell swept up mass) and its 
volume $V_{sh}$

\q\label{eq:rhots}
\rho_{sh}=\frac{M_{sh}}{V_{sh}}=\frac{4}{3}\rho_0 \frac{R_{sh}}{\Delta R_{sh}}.
\nq

For subsequent purposes we define the ratio of gravitational-to-pressure 
force evaluated in the direction along the fragment radius, $R_f$: 

\q\label{eq:f}
f=\frac{\pi G \mu m_H \rho_{sh} R_{f} \Delta R_{sh}}{(3/2)kT_f}.
\nq

\section{Instability in the expanding shell}

We solve numerically Eq. (\ref{eq:mom})-(\ref{eq:energy}) and Eq. 
(\ref{eq:dTb}) for different values of the unperturbed ISM density $n_0$, and 
of the explosion  energy, $E_{SN}$, following the time evolution of the 
shell radius ($R_{sh}$) and velocity ($\dot{R}_{sh}$). 

The unperturbed medium is assumed to be metal free and homogeneous.
The SN progenitor ionizes the medium within its Str\"omgren radius, $R_s =
(3Q_H/4\pi n_0^2 \alpha^2)^{1/3}$, where $Q_H$ is the mass-dependent stellar 
ionizing photon rate and $\alpha$ is the hydrogen recombination rate to levels $\geq 2$.
The radius of the ionized bubble reaches $R_s$ is a recombination time $t_{rec}=(\alpha n_0)^{-1}$;
however, the heated gas will start to freely expand at roughly its sound speed 
($c_s \approx 10$ km/s for an ionized gas) if the dynamical time $R_s/c_s < t_{\star}$, the lifetime
of the star. In this case, by the time of the SN explosion
the ionized region has grown to several tens of pc essentially independent 
of the initial value of $R_s$; inside this region the density and pressure are 
almost constant, thus motivating our constant density approximation.

After the SN
explosion, in the absence of ionizing photons, the gas would recombine on a 
timescale, $t_{rec}$. 
In practice, the gas is kept ionized by the radiation from the shock radiative precursor
and its temperature remains $\approx 10^4$~K until ionizing photons continue to be produced.
This condition is fullfilled as long as the shock velocity is larger than 
$80$~km/s  (Draine \& McKee 1993). It is only a recombination time after such event that
the gas can finally recombine and cool to 300~K.

At each time step we check the supersonic motion
($\dot{R}_{sh}>c_{s,0}$) and instability ($\omega>0$) conditions. 
If both conditions are satisfied we further check that the cooling time is shorter than
the age of the shell, so that the temperature has decreased to $T_f\sim 300$ K.
At this point we calculate the fragment density (Eq. \ref{eq:rhots}), 
mass (Eq. \ref{eq:mfrag}), and the ratio $f$ (Eq. \ref{eq:f}).

\begin{figure}
\centering
\includegraphics[width=0.8\hsize]{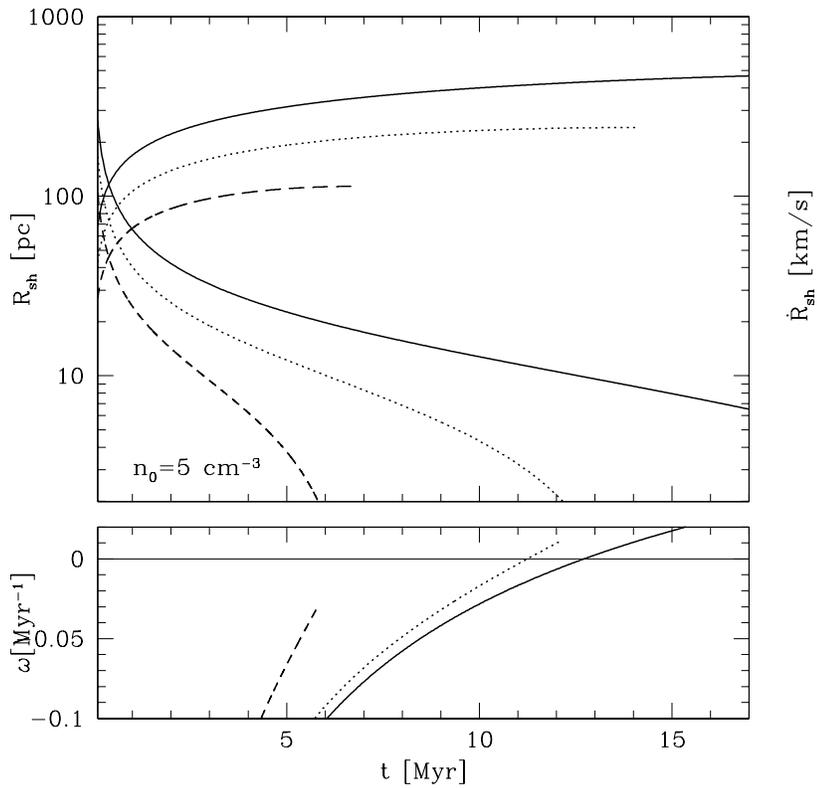}
\caption{{\it Top panel}: shell radius (increasing curves) and velocity as a 
function of time for three SN energies ({\it solid line}: 
$E_{SN}=10^{53}$ erg; {\it dotted line}: $E_{SN}=10^{52}$ erg; 
{\it dashed line}: $E_{SN}=10^{51}$ erg). The ISM density is $n_0=5\;\cm3$. 
{\it Bottom panel}: instability growth rate evolution for the three SN
energies. The curves are plotted until the shock decays into a pressure wave.}
\label{fig:rho2}
\end{figure}

We consider here SN energies in the range $10^{51}-10^{53}$ erg, 
the lower limit being typical of core-collapse Type II SNe (SNII) explosions and the upper limit 
corresponding to the energy released by the most massive 
pair-instability SN ($\SNyy$) explosions (Heger \& Woosley 2002). 
We explore the range of ISM densities $n_0=10^{-2}-10^4\,\cm3$.  
Fig. \ref{fig:rho2} shows the time-evolution of 
the shell radius and velocity for $n_0=5\,\cm3$ 
and three explosion energies. The bottom panel shows 
the corresponding evolution of the instability growth rate $\omega$.
Only the explosions of $\SNyy$ induce shell instability and fragmentation. 

\begin{figure}
\center{{
\includegraphics[width=0.8\hsize]{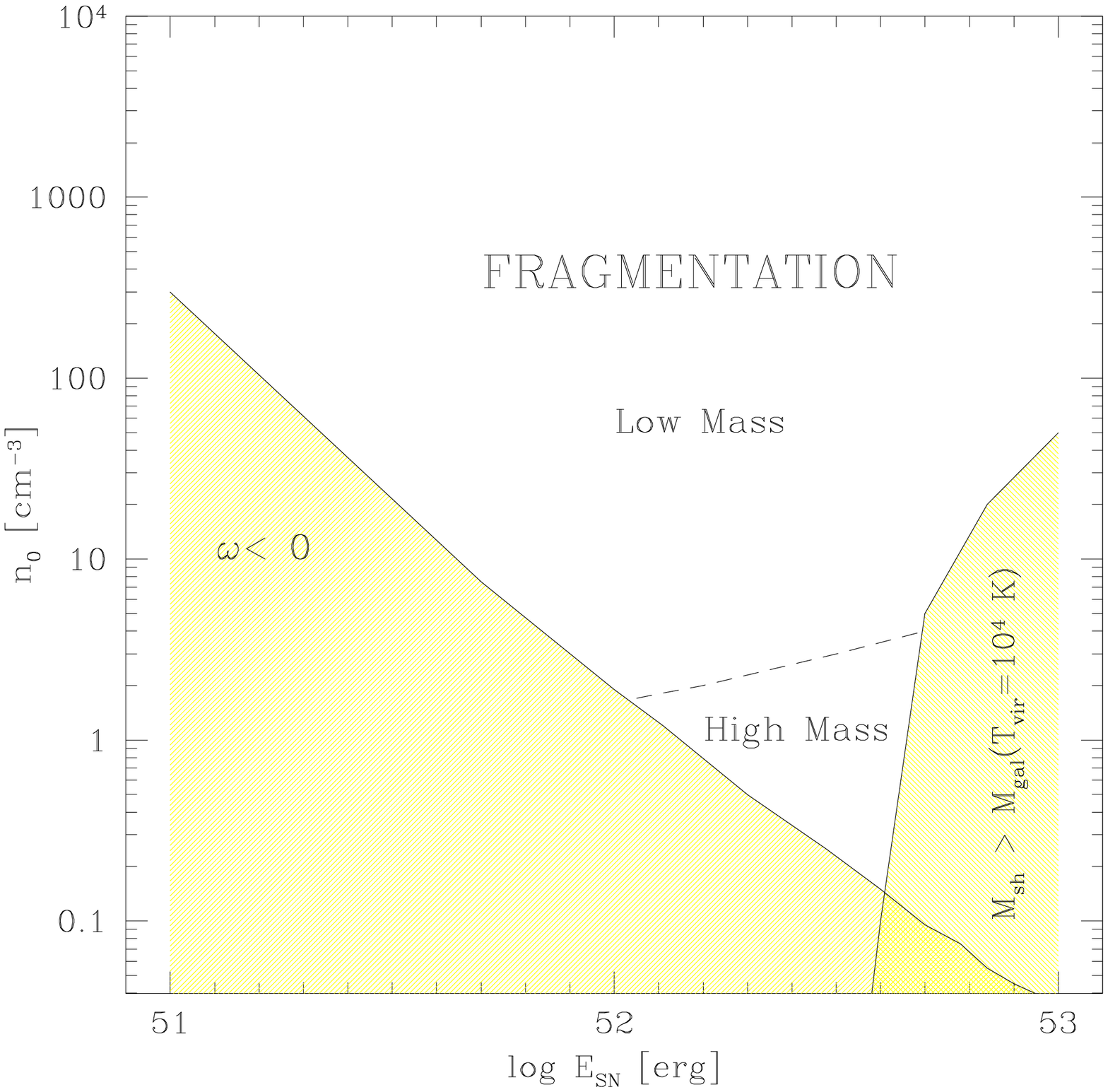}
}}
\caption{Region of the parameter space ($E_{SN}$ -- $n_0$) where fragmentation occurs.
See text for details.}
\label{fig:e_rho}
\end{figure}

Fig. \ref{fig:e_rho} shows the region of the parameter space $E_{SN}-n_0$
where both the conditions $\omega>0$ and $\dot{R}_{sh}>c_{s,0}$ are verified and 
therefore the shell is gravitationally unstable. 
In addition, we also require that the shell mass at the time of instability  
does not exceed the total gas mass of the host galaxy. To exemplify,
in the Figure we assume that the first stars form in halos with mass $M_{gal}$
corresponding to virial temperature
$T_{vir}=10^4$~K. Under these conditions, we see from Fig. \ref{fig:e_rho}
that, if the shell mass does not exceed the baryonic in the galaxy, 
all explosions with energies $E_{SN} \geq 10^{51}$~erg 
can lead to shell fragmentation. However, the minimum ambient density required 
to induce such fragmentation is much larger, $n_0 > 300\;\cm3$, for SNII 
(explosion energy $10^{51}$~erg) than for $\SNyy$, for which the instability can
occur even at lower ambient densities. Hence, pair-instability SNe are more
suitable triggers of induced star formation. 

We conclude that the first pair-instability SNe 
are able to trigger self-propagating 
star formation under a wide range of ambient conditions,
whereas expanding shells created by less energetic `classical' or 
subluminous (as those proposed by Shigeyama 
\& Tsujimoto 1998) SNII        cannot fragment unless the density is 
very high. 

\section{Properties of the fragments}

Let us now explore in more details the properties of the gravitationally unstable
fragments. As it is clear from Eqs. (\ref{eq:lfrag})-(\ref{eq:mfrag}) the 
fragment mass is a function of the explosion energy (through $R_{sh}$ and $\dot{R}_{sh}$),
of the initial density of the medium, $n_0$, and of the density of the 
thin shell, $\rho_{sh}$ (proportional to the shell swept up mass, and thus
function of  $R_{sh}$).
The growth time of the unstable fragments is $\sim \omega^{-1}$.

In all unstable cases, the age of the shell is larger 
than the cooling time and the thin shell 
approximation is valid ($\Delta R_{sh}/R_{sh}<10^{-3}$). 
The instability sets in at 0.2-50 Myr after
the explosion and the typical fragment radius and mass is in the range 
$2-1000$ pc and $10^{4-7}\,\Msun$, respectively 
(the upper limits refer to the lowest ambient density).
The density of the fragments varies between $110\;\cm3$ and $6\times 10^7\;\cm3$;
as we discuss below, this spread has important implications for the mass of 
the second generation of stars formed.
Large values of the 
gravitational-to-pressure force ratio ($f\simeq 8$) are found in all
unstable cases. This is because high shell velocities stretch the 
perturbations, stabilizing the shell (see Eq. \ref{eq:w2}) and  
so the instability conditions are satisfied only when 
the fragments have collected a sufficiently large mass, yielding large values 
of $f$.

NU01 (see their Fig. 6) have shown that primordial filaments fragment into
dense clumps whose masses depend on $n_{sh}$ and $f$. 
For a quasi-equilibrium clump (i.e. $f\simgt1$), fragmentation proceeds 
until the fragment is close to the Bonnor-Ebert mass value corresponding to 
the initial density of the clump\footnote{This is strictly true only for
$n_{sh} \ge n_{cr} = 10^4 \; \cm3$ where $n_{cr}$ is the critical density
which marks the transition from NLTE to LTE regime for H$_2$ cooling (see
discussion in Schneider et al. 2002).} (Palla 2002, Mackey et al. 2003). 
If $f$ is increased to 3, a bifurcation takes place at  $n_{sh}\sim 10^5\;\cm3$. 
For models with $n_{sh}\simgt 10^5\;\cm3$ the minimum fragment mass is 
1-2 $\Msun$, while for $n_{sh}\simlt 10^5\;\cm3$ it is larger than 
$\sim 100\;\Msun$. Further increase of $f$ causes the low mass regime to 
extend down to $n_{sh}\sim 10^4\;\cm3$.

In the low mass regime, the fragmentation proceeds 
down to smaller subclumps searching for an equilibrium between gravitational 
and pressure forces (thus decreasing $f$); at the same time the density of the 
subclumps tends to increase. For $f$ values around 8, 
this equilibrium cannot be reached before the subclump
becomes optically thick to H$_2$ lines. At this stage, the initial clump
has already fragmented into protostellar cores with typical masses 
$\sim 10^{-2}\;\Msun$\footnote{In the case of a strictly metal-free gas the 
minimum mass essentially coincides with the Chandrasekhar mass, i.e. 
$\sim 1-2 \Msun$ (Uehara et al. 1996)}.  
Although the exact value of the resulting stellar mass 
depends on the details of subsequent gas accretion and, possibly, on the
merging of protostellar cores, it is unlikely that these are so efficient
to form a massive star. In fact, the relaxation time corresponding to the
ensemble of $m \sim 10^{-2}\;\Msun$ protostellar cores resulting from the
fragmentation of a $M_f=10^{5}\;\Msun$ clump is $\sim 36$~Gyr.  

As mentioned above, we find $f \approx 8$ for all the unstable case. According
to NU02 (see their Fig. 5a), this $f$ value leads to the preferential formation of low-mass
stars provided the density of the fragment is larger than $\simeq 2 \times 10^3\;\cm3$.
Such condition is verified for all values of the ambient density $n_0$ and explosion
energy $E_{SN}$ except for the small area depicted in Fig. \ref{fig:e_rho}, where instead 
high mass ($M \approx 100\;\Msun$) stars can form as a result of the fragment collapse. 
The formation of high mass stars can only be induced by $\SNyy$, but the range of
suitable ambient densities is small enough that such event can be regarded as unlikely.

\section{Mixing efficiency and metallicity of the fragments}

In order to derive the mean metallicity of the thin shell (and of the
fragments), one should be able to  follow the mixing of SN ejecta with the
shell gas. The standard hydrodynamical response to a sudden release of energy 
implies that the two gases (the swept up matter and the ejecta), after 
crossing their respective shocks find themselves well separated by a contact 
discontinuity. 
However, many mechanisms (e.g. cloud crushing, thermal evaporation, 
hydrodynamical instabilities, effects caused by explosion inside wind-driven
shells and by fragmented ejecta) can disrupt the contact discontinuity
leading to a mixing of the heavy elements into the swept-up matter in the
thin shell (Tenorio-Tagle 1996). A precise description of such physical 
processes is extremely difficult as it requires ultra-high resolution simulations
(de Avillez \& Mac Low 2002; Kifonidis et al. 2003).

The mean metallicity of the shell (or fragment) is

\q
Z_f=f_{mix}\frac{M_{ej}}{M_{sh}}
\nq

\noindent
where $f_{mix}$ is the (unknown) mixing factor of ejected matter in the thin shell and
$M_{ej}$ is the ejected mass in heavy elements. 
For $\SNyy$, 
$M_{ej}\simeq 0.4 M_\star$ (Heger \& Woosley 2002). Even requiring that mixing 
is very efficient (i.e. $f_{mix}=1$), we always find that the mean metallicity 
of the fragments is  $10^{-3.5}\;Z_\odot<Z<10^{-2.6}\;Z_\odot$. 
In more realistic cases, however, mixing is likely to be much more inefficient 
($f_{mix} \ll 1$), so the above value, $10^{-2.6}\;Z_\odot$ is a strict upper limit. 
At face value, if the most iron poor star observed in our Galaxy ([Fe/H] = $-5.3$, Christlieb et al. 2002)
formed through this mechanism, it would imply that $f_{mix} \sim 0.03$. 
Note that in principle zero-metallicity low-mass stars can form
via the proposed mechanism if $f_{mix} = 0$.

\section{Discussion}

We have shown that the explosion of the first SNe 
are able to trigger further star formation.
If the shell mass does not exceed the baryonic in the galaxy,
all explosions with energies $E_{SN} \geq 10^{51}$~erg
can lead to shell fragmentation. However, the minimum ambient density required
to induce such fragmentation is much larger, $n_0 > 300\;\cm3$, for SNII (or subluminous 
supernovae) than for pair-instability ones, which can induce star formation
even at lower ambient densities. 
Fragmenting shells become gravitationally unstable 0.2-50 Myr after the 
explosion, depending on the explosion energy and density of the unperturbed 
medium. 
We have identified the fastest growing mode in the unstable shells and derived
the mass, density and gravitational-to-pressure force ratio, $f$,
of the corresponding fragments. Their
typical mass is in the range $10^{4-7}\;\Msun$ and density in 
the range $110-6\times 10^7\;\cm3$. 
In all cases, we find very large values of the 
gravitational-to-pressure force ratio ($f\simeq 8$), indicating that the formation of 
low-mass, long-living stars is the most likely outcome. A narrow region of the $n_0 - E_{SN}$
parameter space allows the formation of massive stars from fragments originating from 
$\SNyy$ shells. However, the density range is small enough to consider this possibility
as unlikely.  

The mean metallicity of the stars formed through this mechanism depends on the mixing history of the shell which is
largely unknown, as discussed in the previous Section. Depending on the efficiency of this
process and on the combined properties of the explosion energy and ambient density, the 
metallicity of this second generation, low-mass stars can be anywhere in the range
$0 \le Z \le 10^{-2.6}\;Z_\odot$.

The mechanism here proposed is thus able to produce long-living, low-mass extremely metal-poor 
(or even metal-free, if $f_{mix}=0$)
stars that can be found in the Milky Way halo.  These stars can populate the low-mass 
peak of the  bimodal IMF  proposed by NU01 and NU02 or can be
the typical members of the so-called II.5 population (Mackey et al. 2003).
The formation of these second generation stars requires the presence of a 
first generation of massive stars exploding as pair-instability SNe.

So far, only one extremely iron-poor star with [Fe/H]$<-5$ has been identified 
(HE0107-5240, Christlieb et al. 2002). If this result were to be confirmed by the
analysis of the complete volume sampled by the Hamburg/ESO survey,
several implications for the self-propagating star formation mode can be drawn:
the observed lack of [Fe/H]$<-5$ halo stars would imply that mixing of metals in the
unstable shell must occur with moderately high efficiencies ($f_{mix} \simgt 10 \%$).
Alternatively, the mechanism here proposed might not lead to a large number of
observed iron-poor halo stars because: ({\it i}) the typical
mass of this second-generation stars is $\simgt 1-2 \Msun$ so that their lifetimes are not
long enough to be observable as main sequence stars in our Galaxy halo; 
({\it ii}) pair-instability SNe  are
rare because the IMF of the first stars is shaped so that only a small fraction of zero-metallicity 
massive stars form in the mass range of pair-instability SNe, $140 \Msun \simlt M_{\gamma\gamma} \simlt 260 \Msun$. 

Conversely, if more very iron-deficient ([Fe/H]$<-5$) stars were to be identified, the statistics
and properties of these old stellar relics might  lead to important constraints on the dominant
processes which enable low-mass star formation in primordial environments. In particular,  
three viable mechanisms have been proposed for the origin of HE0107-5240 
(Christlieb et al. 2002). Their main difference relies in the interpretation of the observed surface
abundance of C, N  (and O; Christlieb et al. 2004). 
In particular, if one assumes that the observed Fe {\it is not} a good 
indicator of the metallicity of the gas cloud out of which the star formed and that C and N were
already present in the star forming gas (C, N pre-formation scenario), then HE0107-5240 formed with
an initial metallicity of $Z \sim 10^{-2} Z_{\odot}$. Indeed, Umeda \& Nomoto (2003) have shown 
that the abundance pattern of HE0107-5240 is in good agreement with nucleosynthetis yields of a faint
SN explosion of a $\sim 25 \Msun$ zero-metallicity star releasing a kinetic energy of 
$0.3 \times 10^{51}$ erg. If so, due to the small explosion energy, self-propagating star formation
could not have occurred and the star must have formed on a longer timescale from the gas enriched by 
the SN. Alternatively, if one assumes that the observed Fe {\it is} a good 
indicator of the parent cloud metallicity and that C and N were
synthetized in the stellar interior (C, N post-formation scenario)\footnote{These ``post-formation'' 
scenarios require a sensible explanation for the large observed C and N abundances. 
So far, it is still unclear whether efficient mixing during the He-flash can fully 
account for the observed abundances or if it had enough time to operate in HE0107-5240 
(Schlattl et al. 2001, 2002; Siess et al. 2002).}, then HE0107-5240 formed with
an initial metallicity of $Z \sim 10^{-5.1} Z_{\odot}$. Indeed, the observed abundance pattern for
elements heavier than Mg (that cannot be formed in the interior of a $0.8 \Msun$ star such as  
HE0107-5240 and thus retain memory of the nucleosynthesis yields of the pre-enriching star) are well 
reproduced by the predicted yields of a 200 $\Msun$ $\SNyy$. 
In this case, HE0107-5240
could have formed on a short timescale through the self-propagating star formation mechanism or, on a longer
timescale, from the gas enriched by the SN if $20 \%$ of the metals were depleted onto dust 
grains (Schneider et al. 2003; Schneider, Ferrara \& Salvaterra 2003).

\section*{Acknowledgements}
{We thank T.~Abel, F.~Nakamura, M.~Umemura \& K. Omukai for enlightening discussions and useful comments to the paper.}


\begin{thebibliography}{}
\bibitem{} Abel T., Bryan G. L., Norman M. L. 2000, ApJ,
 540, 39
\bibitem{}de Avillez M. A. \& Mac Low M.-M. 2002, ApJ, 581, 1047
\bibitem{}Bromm V., Coppi P. S., Larson R. B. 1999, ApJ, 527, L5
\bibitem{}Bromm V., Ferrara A., Coppi P. S., Larson R. B., 2001 MNRAS, 328, 969
\bibitem{}Bromm V., Coppi P. S., Larson R. B. 2002, ApJ, 564, 23
\bibitem{}Christlieb N., Wisotzki L., Reimers D., Homeier D., Koester D., Heber U. 2001, A\&A, 366, 898
\bibitem{}Christlieb N., Gustafsson B., Korn A. J., Barklem P. S., Beers T. C., Bessell M. S., Karlsson T., Mizuno-Wiedner M. 2004, ApJ, 603, 708
\bibitem{}Christlieb N. et al. 2002, Nature, 419, 904
\bibitem{}Draine B. T. \& McKee C. F. 1993, ARA\&A, 31, 373
\bibitem{}Elmegreen B. G. 1994, ApJ, 427, 384
\bibitem{}Ehlerova S. \& Palous J. 2002, MNRAS, 330, 1022
\bibitem{}Ferrara A. 1998, ApJ, 499, L17
\bibitem{}Heger A. \& Woosley S. E. 2002, ApJ, 567, 532
\bibitem{}Kifonidis K., Plewa T., Janka H.-Th., Mueller E. 2003, A\&A, 408, 621
\bibitem{}Madau P., Ferrara A. \& Rees M. J. 2001, ApJ, 555, 92
\bibitem{}Magliocchetti M., Salvaterra R. \& Ferrara A. 2003, MNRAS, 342, 25
\bibitem{}Mackey J., Bromm V. \& Hernquist L. 2003, ApJ, 586, 1
\bibitem{}Nakamura F. \& Umemura M. 2001, ApJ, 548, 19 (NU01)
\bibitem{}Nakamura F. \& Umemura M. 2002, ApJ, 569, 549 (NU02)
\bibitem{}Omukai K. \& Inutsuka S. 2002, MNRAS, 332, 59
\bibitem{}Omukai K. \& Palla F. 2003,  ApJ, 589, 677
\bibitem{}Ostriker J. P. \& Cowie L. L. 1981, ApJ, 243, L127
\bibitem{}Ostriker J. P. \& McKee C. F. 1988, Rev. Mod. Phys., 60, 1
\bibitem{}Palla F. 2002, in Physics of Star Formation in Galaxies, ed. A. 
Maeder \& G. Maynet (Berlin: Springer), 16
\bibitem{}Salvaterra R. \& Ferrara A. 2003, MNRAS, 339, 973
\bibitem{}Schlattl H., Cassini S., Salaris M., Weiss A. 2001, ApJ, 559, 1082
\bibitem{}Schlattl H., Salaris M., Cassini S., Weiss A. 2002, A\&A, 395, 77 
\bibitem{}Schneider R., Ferrara A., Natarajan P., Omukai K. 2002, ApJ, 571, 30
\bibitem{}Schneider R., Ferrara A., Salvaterra R., Omukai K., Bromm V. 2003, Nature, 422, 869
\bibitem{}Schneider R., Ferrara A. \& Salvaterra R. 2003, MNRAS submitted,
astro-ph/0307087
\bibitem{}Sedov L. I. 1959, Similarity and Dimensional Methods in Mechanics, Academy Press, New York
\bibitem{}Seiss, L., Livio, M. \& Lattanzio, J. 2002, ApJ, 570, 329 
\bibitem{}Shigeyama T. \& Tsujimoto T. 1998, ApJ, 507, L135
\bibitem{}Tenorio-Tagle G. 1996, AJ, 111, 1641
\bibitem{}Uehara H., Susa H., Nishi R., Yamada M., Nakamura T. 1996, ApJ, 473, L95
\bibitem{}Umeda H. \& Nomoto K. 2003, Nature, 422, 871
\bibitem{}W\"{u}nsch R. \& Palous J. 2001, A\&A, 374, 746

\end{thebibliography}
\end{document}